\documentclass[useAMS, referee]{mn2e}
\input{psfig.sty}
\usepackage{graphicx}
\usepackage{times}

\begin{document}
\title{Distributions of the Hardness Ratio of short and Long Gamma-Ray Bursts in Different Time
Intervals within the First 2 Seconds}
%\footnote{send offprint request to: Yun-Ming Dong}}

\date{2004 Sept. 8}
\pubyear{????} \volume{????}

\pagerange{2} \onecolumn

\author[Y.M. Dong et al.]
       {Yun-Ming Dong$^{1,2}$, Yi-Ping Qin$^{1,3}$
 \\
        $^1$ National Astronomical Observatories/Yunnan Observatory,
         Chinese Academy of Sciences, P. O. Box
110, Kunming 650011, China; \\dongym@ynao.ac.cn\\
        $^2$ Graduate School of The Chinese Academy of
Sciences
 \\
        $^3$ Physics Department, Guangxi University, Nanning,
Guangxi 530004, P. R. China \\
 }

\date{Accepted ????.
     Received ????}

%\pagerange{\pageref{firstpage}--\pageref{lastpage}} \pubyear{2004}

\maketitle

\label{firstpage}

\begin{abstract}

In the present paper, we investigated the distribution of hardness
ratio (HR) for short and long gamma-ray bursts (GRBs) in different
time scales for the first two seconds. After including and
subtracting the background count, we performed a
Kolmogorov--Smirnov (K-S) test to the HR distributions of the two
classes of GRBs in each time interval. We obtained from our
analysis, the HR distributions of the two classes of bursts are
obviously different. The result indicates that the two kinds of
bursts probably originate from the different mechanisms or have
different central engines. In addition, we found that the hardness
ratio of short bursts within the time interval of 0$-$0.96 seconds
changes hard-to-soft, on the other hand long bursts do not. The
two kinds of bursts have different characteristics in the first 2
seconds which might be associated with different physical
mechanisms.

% We investigate distributions of the hardness ratio (HR) of
%  short and long gamma--ray bursts (GRBs) in different time intervals
%  within the first two seconds when including and subtracting the
%  background count, respectively, and perform a Kolmogorov--Smirnov (K-S)
%  test to the HR distributions of the two classes of GRBs in each
%  time interval. Obtained in our analysis,
%  K-S test to the distributions is very small, showing that HR distributions of the
%  two classes of bursts are obviously different. This indicates that the two kinds of bursts
%  probably originate from different mechanisms or have different central engines.
%  In addition, we find that the hardness ratio of short bursts within the time interval
%  of 0--0.96 seconds changes following hard--to--soft, while that of long
%  bursts does not. The two kind of bursts have different characters in the
%  first 2 seconds which might be associated with different physical processes.

\end{abstract}

\begin{keywords}
gamma rays: bursts-- methods: statistical
\end{keywords}

\section{Introduction}

Based on the bimodal time interval distribution, gamma--ray bursts
(GRBs) can be divided into two subclasses with T90$=$2s, first one
is the short burst with a mean T90 $\sim$ 0.3s and observed event
rate is $\sim$ 1/3 of the long bursts and the second one is long
burst class with T90 $\sim$ 20s (Kouveliotou et al. 1993; Norris
et al. 2000). Statistical studies revealed that these two classes
have different distributions of the hardness ratio (short bursts
being harder), pulse width, separation time scale, number of
pulses per bursts, different anti--correlations between the
spectral hardness and duration. It suggests that the two classes
might be intrinsically different (Qin et al. 2000, 2001; hurley et
al. 1992; Kouveliotou et al. 1993; fishman \& meegan 1995, Norris
et al. 2000; Nakar \& Piran 2002). A generally accepted scenario
is that short bursts are likely to be produced by the merger of
compact objects while the core collapse of massive stars is likely
to give rise to long bursts (see, e.g., Zhang \&
M$\acute{e}$saz$\acute{a}$ros, 2003; Piran, 2004).

%Based on the bimodal duration distribution, gamma--ray bursts
%(GRBs) have been divided into two subclasses with $T90=2s$: one is
%the short burst class with a mean T90 of $\sim$ 0.3s, which
%observed event rate is the $\sim$ 1/3 of that of long bursts and
%the other is long burst class with a mean T90 of $\sim$ 20s
%(Kouveliotou et al. 1993; Norris et al. 2000). Statistical studies
%revealed that the two classes have different distributions of the
%hardness ratio (short bursts being harder), pulse width,
%separation time scale, and number of pulses per bursts, and have
%different anti--correlations between the spectral hardness and
%duration, suggesting that the two classes might be intrinsically
%different (Qin et al. 2000, 2001; hurley et al. 1992; Kouveliotou
%et al. 1993; fishman \& meegan 1995, Norris et al. 2000; Nakar \&
%Piran 2002). A generally accepted scenario is that short bursts
%are likely to be produced by the merger of compact objects while
%the core collapse of massive stars is likely to give rise to long
%bursts (see, e.g., Zhang \& M$\acute{e}$saz$\acute{a}$ros, 2003;
%Piran, 2004).

However, there are some evidences indicating that the two classes
might be originated from the same physical process. Schmidt (2001)
suggested that both short and long bursts have a similar
luminosity. Lamb et al. (2002) found that $<V/V_{max}>$, the
angular distribution, the energy dependence of the duration and
the hard$-$to$-$soft spectral evolution of short bursts are
similar to those of long bursts. Recently, Ghirlanda, Ghisellini
\& Celotti (2004) have shown that the emission properties of short
bursts are similar to that of the first 2 seconds for long events
and concluded that the central engine of long and short bursts is
the same, only working for a longer time for long GRBs. Based on
this work, Yamazaki, Ioka \& Nakamura (2004) proposed a unified
model of short and long GRBs and suggested that the jet of GRB
consists of multiple subjets or subshells, where the multiplicity
of the subjets along the line of sight $n_{s}$ is an important
parameter. They showed that if $n_{s}$ is large ($\gg$ 1), the
event looks like a long GRB, while if $n_{s}$ is small ($\sim$ 1),
the event looks like a short GRB.

%However, there are some evidences indicating that the two classes
%might originate from the same physical process. Schmidt (2001)
%suggested that both short and long bursts have a similar
%luminosity. Lamb et al. (2002) found that $<V/V_{max}>$, the
%angular distribution, the energy dependence of the duration, and
%the hard-to-soft spectral evolution of short bursts are similar to
%those of long bursts. Recently, Ghirlanda, Ghisellini \& Celotti
%(2004) showed that the emission properties of short bursts are
%similar to that of the first 2 seconds of long events and they
%suggested that the central engine of long and short bursts is the
%same, just working for a longer time for long GRBs. Based on this
%work, Yamazaki, Ioka \& Nakamura (2004) proposed a unified model
%of short and long GRBs, suggesting that the jet of GRB consists of
%multiple subjets or subshells, where the multiplicity of the
%subjets along a line of sight $n_{s}$ is an important parameter.
%They showed that if $n_{s}$ is large ($\gg$ 1), the event looks
%like a long GRB, while if $n_{s}$ is small ($\sim$ 1), the event
%looks like a short GRB.

Thus, it is still unclear whether the short and the long bursts
are intrinsically the same. To find the mechanism of the central
engine of GRBs, it is necessary to solve this question with
focused effort. Based on the analysis of Ghirlanda et al. (2004),
it is expected that the first 2 second behavior of short and long
bursts would be the same. If so, the hardness ratio distributions
of the two classes should share the same character within this
period. This will be checked in the following.

%Thus, it is still unclear whether the short and the long bursts
%are intrinsically the same. To find the mechanism of the central
%engine of GRBs, it is necessary to make clear this question. Based
%on the analysis of Ghirlanda et al (2004), it is expected that the
%first 2 second behaviors of short and long bursts would be the
%same. If so, the hardness ratio distributions of the two classes
%would share the same character within this period. This will be
%checked in the following.

Distributions of the hardness ratio of short and long GRBs for
each 64 ms time interval within the first 2 seconds of bursts are
presented in section 2. In section 3 we present the results of the
K-S test to the distributions of the two kinds of bursts, where
some typical statistical values of the distributions are also
provided. A brief discussion is presented in section 4.

%Distributions of the hardness ratio of short and long GRBs for
%each 64 ms time interval within the first 2 seconds of bursts are
%presented in section 2. While in section 3 we present the results
%of the K-S test to the distributions of the two kinds of bursts,
%where some typical statistical values of the distributions are
%also provided. A brief discussion is presented in section 4.

\section{Distributions of the Hardness
Ratio of Short and Long GRBs within the First 2 seconds}

There are 2041 GRBs with T90, including 500 short GRBs and 1541
long GRBs, available in the current BATSE burst catalog. In this
catalogue observations made during April 21, 1991 to May 26, 2000
by BATSE with 64 ms temporal resolution and four-channel spectral
resolution were presented. We found that this catalogue has 462
short bursts and 1428 long bursts.

%There are 2041 GRBs with T90, including 500 short GRBs and 1541
%long GRBs, available in the current BATSE burst catalog, where 64
%ms temporal resolution and four-channel spectral resolution GRB
%data observed by BATSE from 1991 April 21 to 2000 May 26 are
%presented. We find from this catalogue 462 short bursts and 1428
%long bursts with their 64 ms data being available.

In this paper we mainly focused our attention on the question,
whether the two classes have the same hardness ratio distribution
within the first 2 seconds? The hardness ratio is defined as $HR_n
\equiv count_3/count_2$, where $count_2$ and $count_3$ are the
counts of the second channel and the third channel within the 64
ms time interval respectively, $n$ represents the $n$th 64 ms time
interval after the trigger time. We considered two different
situations, the hardness ratio is calculated with the original
data which contain the background counts (case 1), and the
hardness ratio is calculated when the background counts are
subtracted (case 2).

%In this paper we focus our attention mainly on the question
%whether the two classes have the same hardness ratio distribution
%within the first 2 seconds. The hardness ratio is defined as $HR_n
%\equiv count_3/count_2$, where $count_2$ and $count_3$ are the
%counts of the second and third channels, respectively, within the
%64 ms time interval, and $n$ represents the $n$th 64 ms time
%interval after the trigger time. We consider two different
%situations. One is that the hardness ratio is calculated with the
%original data which contain the background counts (case 1) and the
%other is that the hardness ratio is calculated when the background
%counts are subtracted (case 2).

In case 1, we have two ways for calculating the hardness ratio for
short bursts: one is to calculate the hardness ratio for any of
the 64 ms interval from the trigger time to the end of T90 (in
this way some sources will be missed within the time interval
between the end of T90 and the end of the first 2 seconds due to
their short durations); the other way is to calculate the ratio
for all 64 ms intervals within the first 2 seconds of bursts (in
this way, no sources will be missed in any interval within the
first 2 seconds). For long bursts, as their duration is larger
than 2 seconds, we adopted the second approach applied for the
short bursts. We therefore, obtained three distributions of the
hardness ratio for any of the 64 ms interval concerned which we
call distributions 1, 2 and 3, respectively.

%In case 1, we have two ways for calculating the hardness ratio for
%short bursts: one is to calculate the hardness ratio for any of
%the 64 ms interval from the trigger time to the end of T90 (thus
%there will be some sources missed within the time interval between
%the end of T90 and the end of the first 2 seconds due to their
%short durations); the other is to calculate the ratio for all 64
%ms intervals within the first 2 seconds of bursts (in this
%situation, there will be no sources missed in any interval within
%the first 2 seconds). For long bursts, as their duration is larger
%than 2 seconds, the second approach adopted in the case of short
%bursts is applied. We therefore obtain three distributions of the
%hardness ratio in this case for any of the 64 ms interval
%concerned which we call distributions 1, 2 and 3, respectively.

In case 2, for each source, we assumed its signal data covers the
range of $t_{min}\leq t\leq t_{max}$, where
$t_{max}-t_{min}=2T_{90}$, and $t_{min}$ is at $T_{90}/2$ previous
to the start of $T_{90}$. If $t_{min}$ is previous to the start of
the data, we then assigned $t_{min}$ to be the start of trigger
time. Data beyond this range will be taken to find the fit for the
background, which will be fitted with a liner function. This
background fit will then be applied to the signal interval and be
taken as the background count rate there. We used the data in the
first 2 seconds which subtract the corresponding background counts
to calculate the corresponding hardness ratio. For short bursts,
we calculated the hardness ratio from the trigger time to the end
of T90 and obtained the corresponding distribution (distribution
4). For long bursts we calculated the hardness ratio in the first
2s and got the corresponding distribution (distribution 5).

%In case 2, for each source, we assume its signal data covers the
%range of $t_{min}\leq t\leq t_{max}$, where
%$t_{max}-t_{min}=2T_{90}$, and $t_{min}$ is at $T_{90}/2$ previous
%to the start of $T_{90}$. If $t_{min}$ is previous to the start of
%the data, we then assign $t_{min}$ to be the start of trigger
%time. Data beyond this range will be taken to find the fit of the
%background, which will be fitted with a liner function. This
%background fit will then be applied to the signal interval and be
%taken as the background count rate there. We use the data in the
%first 2 seconds which subtract the corresponding background counts
%to calculate the corresponding hardness ratio. For short bursts,
%we calculate the hardness ratio from the trigger time to the end
%of T90 and obtain the corresponding distribution (distribution 4).
%For long bursts we calculate the hardness ratio in the first 2s
%and get the corresponding distribution (distribution 5).

The three distributions of case 1 are presented in Fig. 1 and the
two distributions of case 2 are presented in Fig. 2. There are 27
plots in total for each of the two figures, which represent
intervals 0--0.064 s (hr$_{1}$), 0.064--0.128s (hr$_{2}$), ...,
and 1.664--1.728s (hr$_{27}$), respectively. As the number of the
hardness ratio data after the 1.728s is very small for short
bursts, only 27 plots are presented.

%The three distributions of case 1 are presented in Fig. 1 and the
%two distributions of case 2 are presented in Fig. 2. There are 27
%pictures in total for each of the two figures, which represent
%intervals 0--0.064 s (hr$_{1}$), 0.064--0.128s (hr$_{2}$), ...,
%and 1.664--1.728s (hr$_{27}$), respectively. As the number of the
%hardness ratio data after the 1.728s is very small for short
%bursts, only 27 pictures are presented.

\begin{figure}

 %\vspace{5.5cm}
  \includegraphics[width=6in,angle=0]{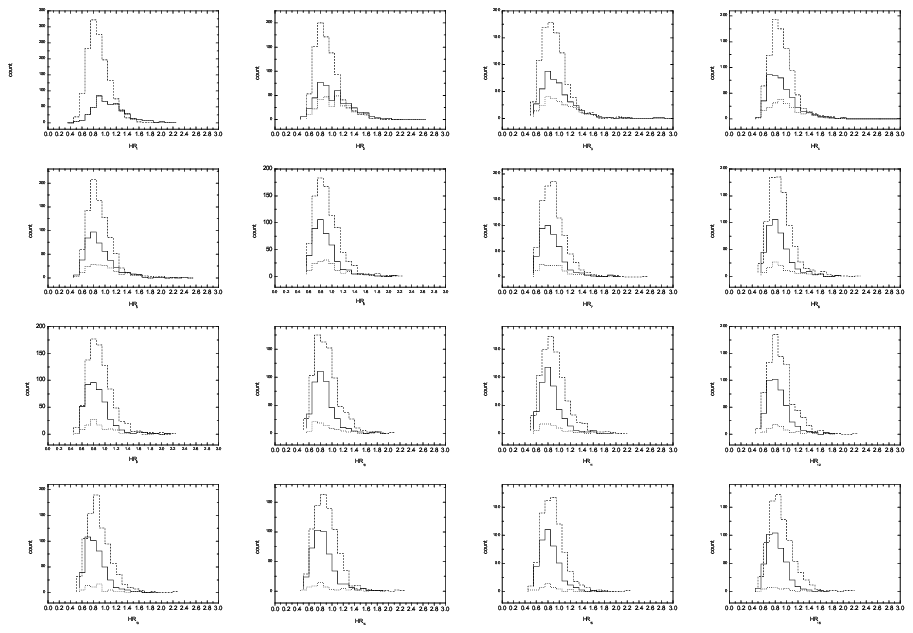}
   \includegraphics[width=6in,angle=0]{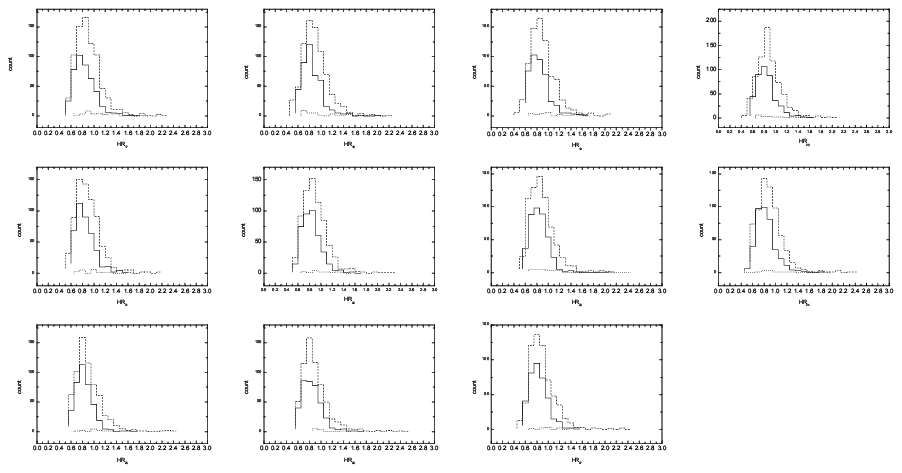}

   \caption{Distributions of the hardness ratio associated with time for short
   and long bursts in case 1. Dashed lines represent long bursts
(distributions 3), solid lines represent short bursts calculated
within the first 2 seconds (distributions 2), and dotted lines
stand for short bursts calculated within T90 (distributions 1),
respectively}
  \label{Fig1}
\end{figure}
\begin{figure}

 %\vspace{5.5cm}
  \includegraphics[width=6in,angle=0]{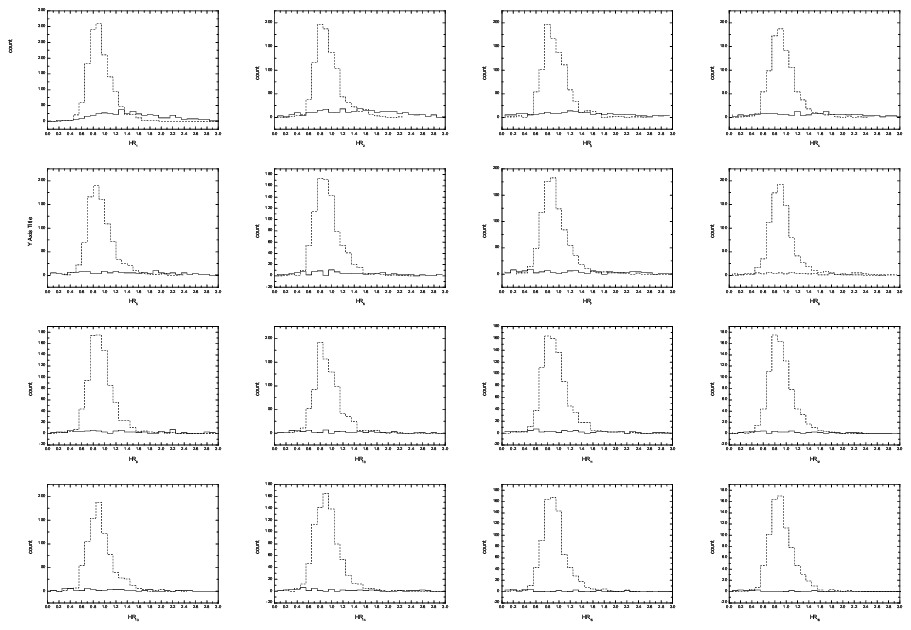}
   \includegraphics[width=6in,angle=0]{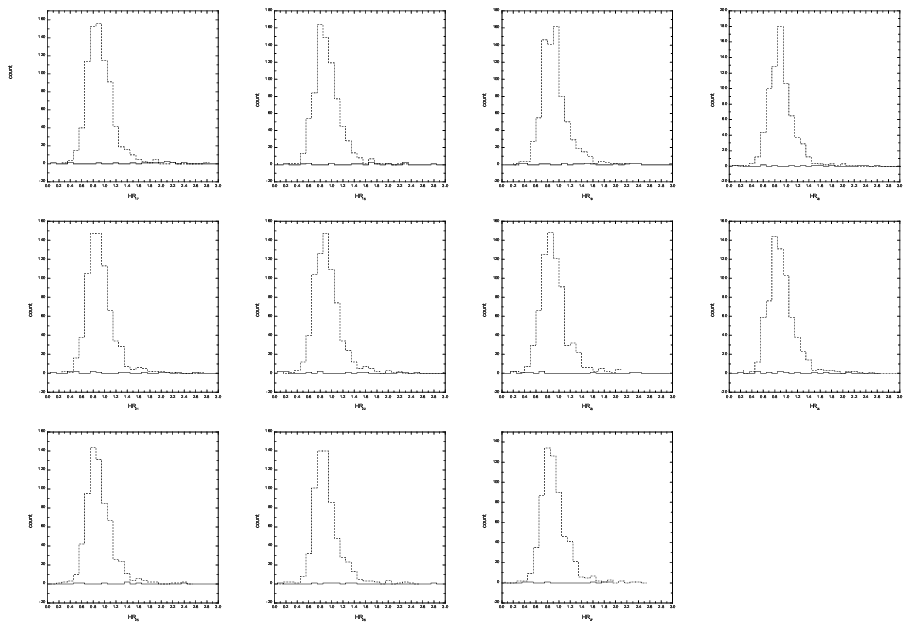}

   \caption{Distributions of the hardness ratio associated with time for short
   and long bursts in case 2. Dashed lines represent long bursts
(distributions 5), and solid lines denote short bursts
(distributions 4), respectively}
  \label{Fig2}
\end{figure}

\section{The K-S test to and the Median value of the Hardness Ratio Distributions}

In order to check if the behavior of the two classes of GRBs are
the same within the first 2 seconds, we performed a K-S test to
the hardness ratio distributions mentioned above. The results are
listed in Table 1, where median values of the distributions are
also presented. As shown in the table, probability 1 represents
the probability associated with the K-S test to the distributions
1 and 3, probability 2 denotes the probability obtained from the
K-S test to the distributions 2 and 3, while probability 3
describes the probability produced by the K-S test to the
distributions 4 and 5. We have also reported in the table, median
1, 2, 3, 4, and 5 which are the median values of the distributions
1, 2, 3, 4, and 5, respectively.
%In order to check if the behaviors of the two classes of GRBs are
%the same within the first 2 seconds, we perform a K-S test to the
%hardness ratio distributions mentioned above. The results are
%listed in Table 1, where median values of the distributions are
%also presented. As shown in the table, probability 1 represents
%the probability associated with the K-S test to the distributions
%1 and 3, and probability 2 denotes the probability obtained from
%the K-S test to the distributions 2 and 3, while probability 3
%describes the probability produced by the K-S test to the
%distributions 4 and 5. Also in the table, median 1, 2, 3, 4, and 5
%are the median values of the distributions 1, 2, 3, 4, and 5,
%respectively.
\begin{table*}
\caption{Probability of the K-S test to the hardness ratio
distributions of the two classes of GRBs} \centering
\begin{tabular}{lllllllll}
\hline\hline
 intervals &probability 1&probability 2&probability 3& median 1& median 2 & median 3& median 4& median 5\\
      \hline\\
    1 & 0.1938E-29 & 0.1938E-29 & 0.0000E+00 &  1.094 &  1.094 &  0.912&1.503&   0.945\\
    2 & 0.3832E-17 & 0.6548E-12  &0.0000E+00 & 1.107  & 1.028  & 0.929&1.565&   0.949\\
    3 & 0.6243E-07 & 0.1775E-02  &0.3559E-37 & 1.018  & 0.964  &0.934&1.435 &  0.959\\
    4 &0.2438E-02 & 0.8775E-01 & 0.5517E-31& 1.000 & 0.924 & 0.936&1.477 &0.962\\
    5  &0.9181E-03& 0.2878E+00 &0.3965E-23&  0.997 &  0.904  &0.923&1.327  & 0.950\\
    6  &0.2084E+00 & 0.9833E-02 &0.3041E-16&  0.952 & 0.887  & 0.926&1.142 &  0.953\\
    7 & 0.7741E-01&  0.1125E-01& 0.2415E-15&  0.965&   0.885 &  0.920&1.248 &  0.945\\
    8  &0.8181E-01&  0.6498E-04 & 0.9417E-13& 0.918 &  0.878 &  0.922&1.163  & 0.948\\
    9 & 0.3096E+00&  0.5504E-03 & 0.2510E-09 &0.914  &0.884 &  0.930&1.015 &  0.960\\
   10 & 0.6099E+00&  0.2328E-04 &0.3260E-12&  0.913 &  0.873  & 0.919&0.814 &  0.948\\
   11 & 0.8810E+00 & 0.6582E-06 & 0.6947E-08& 0.912 &  0.870 &  0.926&0.977 &  0.954\\
   12 &0.8999E+00&  0.4512E-03 &0.2112E-06&  0.909 &  0.867 &  0.909&0.955  & 0.938\\
   13 & 0.5467E+00 & 0.5624E-04 & 0.1175E-06& 0.909 &  0.867 & 0.916&0.800 &  0.944\\
   14  &0.2414E+00 & 0.1957E-06  &0.8912E-09& 0.870 &  0.869 &  0.924&0.733  & 0.957\\
   15  &0.9558E-01 & 0.1222E-05 &0.1985E-10   &   0.862  & 0.862 &  0.921&0.543 & 0.956\\
   16  &0.7092E+00 & 0.5563E-05  &0.1015E-07   & 0.897 &  0.853  & 0.918&0.615 &  0.945\\
   17 & 0.3297E-01& 0.3521E-03 &0.3483E-06   &  1.049 &  0.865  & 0.916&1.653 &  0.948\\
   18 & 0.1441E+00 & 0.3280E-05 &0.3488E-06   & 0.903 &  0.862  & 0.916&1.631  & 0.949\\
   19 & 0.6253E-01&  0.6565E-04&0.1631E-04   &   1.014  & 0.864  & 0.910&1.520 &  0.951\\
   20  &0.4470E+00 & 0.5224E-04&0.1564E-02   &   0.953 &  0.862 &  0.910&0.820  & 0.936\\
   21 & 0.1915E+00 & 0.1176E-02&0.1496E-01   &   1.020 & 0.866 &  0.902&0.877 &  0.936\\
   22 & 0.3636E+00 & 0.4090E-04 &0.9226E-02   &  0.932 &  0.870&   0.917&0.872 & 0.946\\
   23  &0.7136E+00&  0.1043E-02 &0.3459E-03   &  0.957 & 0.872  & 0.898&0.561  & 0.931\\
   24 & 0.4024E+00 &0.2717E-04  &0.3804E+00   & 0.958 &  0.860  & 0.908&0.981 &  0.937\\
   25 & 0.3405E-01 & 0.3798E-04 &0.1515E-01   &  1.021  & 0.857  & 0.895&1.003 &  0.926\\
   26  &0.4801E-01 & 0.1199E-01 &0.2888E+00    &  0.958  & 0.871  &0.890&0.978 &  0.921\\
   27 &0.3061E-01 &0.3113E-02 & 0.6111E-02   & 1.067  & 0.862 &  0.900&0.550   &0.930\\

\hline
\end{tabular}

%\end{center}
\end{table*}

\begin{figure}
\centering
%  \vspace{5.5cm}
  \includegraphics[width=3.5in,angle=0]{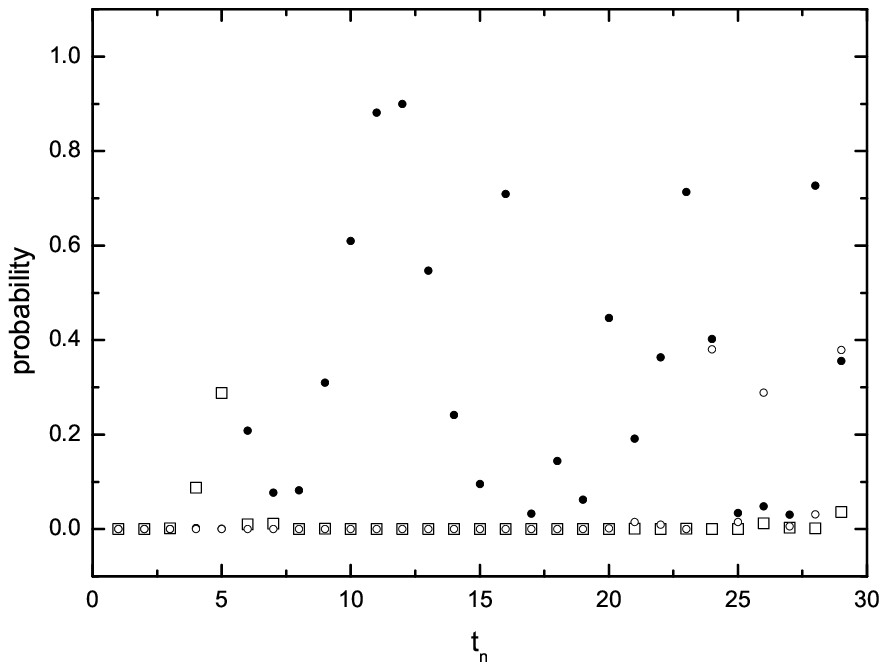}

   \caption{Probability of the K-S test associated with different 64 ms time intervals within the first 2 seconds.
   Filled circles represent the probability obtained from the K-S test to distributions 1 and 3,
   and open squares stand for the probability associated with the K-S test to distributions 2 and 3, and open
   circles denote the probability produced by the K-S test to distributions 4 and 5,
   respectively.}
  \label{Fig3}
\end{figure}

From Table 1 and Fig. 3, we find that probabilities associated
with the K-S test to distributions 1 and 3 vary significantly. In
the first five time intervals, the probability is very small (it
is very close to zero), indicating that during this period the
hardness ratio distributions of the two classes of GRBs are
unlikely to arise from a same parent population. From the 8th time
interval to the 12th time interval, the probability rises
monotonously and reaches its maximum $\sim$ 0.9. Within the 12th
and 15th time intervals, it drops monotonously and reaches its
rock bottom at the 15th time interval. After the 15th time
interval, the probability curve seems to arise from fluctuation
which we believe to be due to the relatively small number of short
bursts (in fact, after the 15th time interval, the number of short
bursts is $\sim 33$ which is much smaller than that of long
bursts). The probability of the K-S test associated with
distributions 2 and 3 is significantly small ($\sim 0$) which
indicates that in this case the hardness ratio distributions of
the two GRB classes are unlikely to arise from the same parent
population. Note that, after T90, the hardness ratio distribution
of short bursts reflects the background count ratio rather than
the signal count ratio. It is, therefore, reasonable that the
probability associated with distributions 2 and 3 is $\sim 0$. For
distributions 4 and 5, for which the background count has been
subtracted, the probability of the K-S test is very small (see
Table 1 and Fig. 3), which suggest that the hardness ratio
distributions of the two classes of the GRBs are unlikely to arise
from the same parent population and therefore they are likely to
originate from different physical processes.

For distributions 1 and 3, in the 11th and 12th time intervals the
probability of the K-S test is very large (close to 1). Can we
draw a conclusion that the two classes originated from a same
physical process in these time intervals? The answer is no. If
short and long GRBs originate from a same physical process and
have a same central engine as suggested in some previous works
(see, e.g., Ghirlanda et al. 2004; Yamazaki et al. 2004), they
should have the same hardness ratio distributions within the first
2 seconds. But the fact is that, in the first 5 time intervals the
K-S probability is almost $\sim 0$ and it is $\sim 1$ only in the
11th and 12th time intervals, which is in confliction with the
scenario that short and long GRBs originate from the same physical
process. Note that a character of the development of the GRB
spectrum is hard-to-soft. Since the hardness ratio of short bursts
varies enormously, it is expected that within some time interval,
the hardness ratio of short bursts can become the same as that of
long bursts, and this we believe to be the reason why the
possibility of the K-S test of the two hardness ratio
distributions in the 11th and 12th time intervals is as large as
$\sim 1$. This is supported by the median value of the hardness
ratio distributions shown in fig. 5, where the median values of
short and long bursts from the 8th to 12th time intervals are
almost the same.

%For distributions 1 and 3, in the 11th and 12th time intervals the
%probability of the K-S test is very large (close to 1). Can we
%draw a conclusion that the two classes originate from a same
%physical process in these time intervals? The answer is no. If
%short and long GRBs originate from a same physical process and
%have a same central engine as suggested in some previous works
%(see, e.g., Ghirlanda et al. 2004; Yamazaki et al. 2004), they
%should have the same hardness ratio distributions within the first
%2 seconds. But the fact is that, in the first 5 time intervals the
%K-S probability is almost $\sim 0$ and it is $\sim 1$ only in the
%11th and 12th time intervals, which is in confliction with the
%scenario that short and long GRBs originate from the same physical
%process. Note that a character of the development of the GRB
%spectrum is hard-to-soft. Since the hardness ratio of short bursts
%varies enormously, it is expected that within some time interval,
%the hardness ratio of short bursts can become the same as that of
%long bursts, and this we believe to be the reason why the
%possibility of the K-S test of the two hardness ratio
%distributions in the 11th and 12th time intervals is as large as
%$\sim 1$. This is supported by the median value of the hardness
%ratio distributions shown in fig. 5, where the median values of
%short and long bursts from the 8th to 12th time intervals are
%almost the same.

However, for distributions 4 and 5 for which the background count
is subtracted, the probability of the K-S test in all the 64 ms
time intervals concerned is almost 0. If the method of subtracting
the background count is reasonable, we can take the data
subtracting the background count as the true signal data of the
bursts. In this situation, it would be natural to conclude from
the analysis that the two GRB classes are unlikely to be
intrinsically the same. This can be convinced when one makes a sum
of counts of the signal data within the first 2 seconds and
calculates the corresponding hardness ratios and then performs a
K-S test to the hardness ratio distributions of the two classes of
bursts, as is shown in the following.

In addition, besides making a sum of counts within the first 2
seconds for long bursts, we also calculate the sum of counts
ranging from the trigger time to the end of T90 for short bursts,
in both the cases of including and subtracting the background
counts. We then calculate the corresponding hardness ratios. The
hardness ratio distributions of short and long bursts in the two
cases are presented in Fig. 4. The probabilities of the K-S test
to the hardness ratio distributions of short and long bursts in
the two cases are 1.1153706E-13 (including the background count)
and 2.4111853E-27 (subtracting the background count),
respectively. They are so small that the two kinds of bursts are
unlikely to be associated with the same physical process and to
have a same central engine.

\begin{figure}
\centering
%  \vspace{5.5cm}
  \includegraphics[width=3.5in,angle=0]{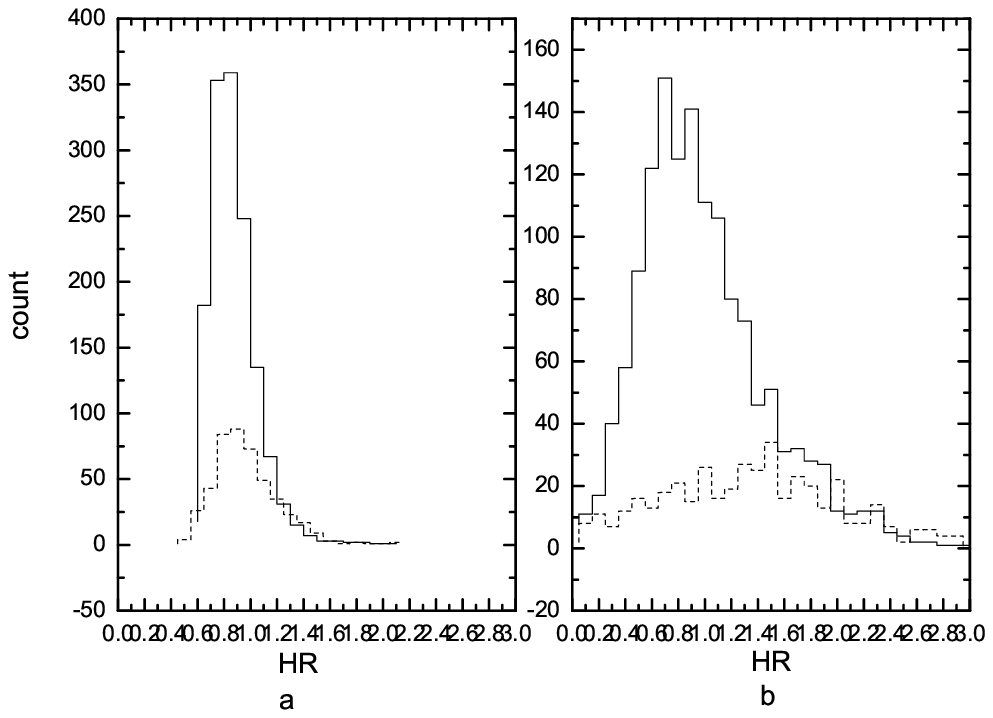}

   \caption{Distributions of the hardness ratio within the first two seconds for
    short and long bursts in cases 1 (denoted a) and 2 (denoted b), where solid
    lines represent the distribution of long bursts and dashed lines
stand for that of short bursts.}
  \label{Fig4}
\end{figure}

In Fig. 5 we have shown all the median values of distributions 1,
2, 3, 4, and 5. We find that for long bursts, the median values do
not show a fluctuation and keep almost the same. For short bursts,
within the period before the 15th, 64 ms time interval (the range
of 0--0.96s), distributions 1, 2, and 4 show a same trend:
monotonously dropping. After this time interval, the median of
distribution 4 exhibits an obvious character of fluctuation, while
for distribution 1 the character is insignificant. Meanwhile, the
median of distribution 2 keeps to be the same. One can conclude
from this analysis that short and long bursts are intrinsically
different: the hardness ratio of long bursts does not obviously
develop with time, while that of short bursts evolves
significantly and it becomes much softer when time goes on. In
addition, we find that, during the beginning phase of bursts, the
hardness ratio of short GRBs is larger than that of long ones,
which is accordant with the result that short bursts are typically
harder (Dezalay et al. 1996).

\begin{figure}
\centering
%  \vspace{5.5cm}
  \includegraphics[width=3.5in,angle=0]{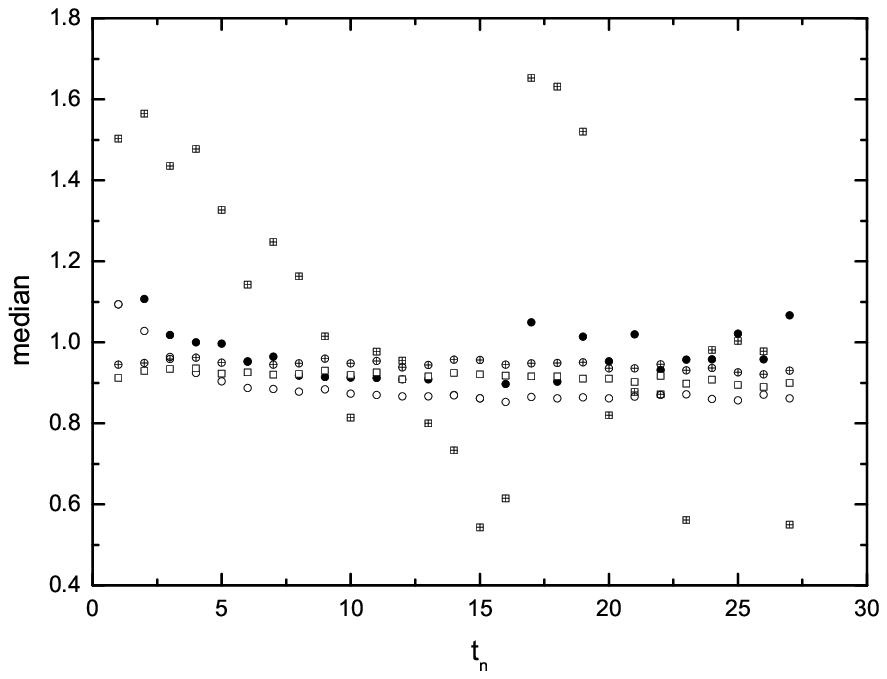}

   \caption{Median of distributions 1, 2, 3, 4 and 5 in
    different 64 ms intervals within the first 2 seconds. The filled circle, open
    circle, open square, open square plus cross, and open circle plus cross
   represent distributions 1, 2, 3, 4 and 5, respectively.}
  \label{Fig5}
\end{figure}

\section{Discussion and conclusions}

Distributions of the hardness ratio in different 64 ms time
intervals within the first 2 seconds for short and long bursts are
studied in the cases of including (case 1) and subtracting (case
2) the background count. The main aim of this work is to find in
the first two seconds whether the hardness ratio of short and long
bursts evolves and the two classes of GRBs arise from a same
physical process and have a same central engine. From Figs. 1 and
2, no significant evolution of the hardness ratio can be detected.
In order to find an evolutionary clue, we investigated the median
value of these distributions, as shown in Fig. 5. From Fig. 5 we
found that the median value of short bursts monotonously drops
with time from 0 to 0.96s, and that of long bursts does not show a
significant evolution. This implies that for short bursts their
spectra evolve significantly with a character of hard-to-soft
within the first 2 seconds while for long bursts their spectra do
not show an obvious change during this period.

A K-S test to hardness ratio distributions of short and long
bursts in both cases of including and subtracting the background
count for the same 64ms time intervals shows that for the majority
of the time intervals concerned the probability is very small
(close to zero). This indicates that in these time intervals the
mechanisms of the two types of bursts are unlikely to be
intrinsically the same. Making a sum of counts over the whole
first 2 seconds and performing a K-S test to the corresponding
hardness ratio distributions we can also reach the same
conclusion. Our analysis suggests that short and long GRBs
probably originate from different physical processes or have
different central engines and they are intrinsically different and
distinct subclasses.

We would like to address our great thanks to Alok C. Gupta for
correcting the English of this paper. We would also like to thank
anonymous referee for his or her several incisive and substantive
comments and suggestions, which have allowed us to improve this
revised paper greatly. This work was Supported by the Special
Funds for Major State Basic Research ("973") Projects and the
National Natural Science Foundation of China (grants 10273019).

\label{lastpage}

\end{document}